\documentclass{article}
\usepackage{amsmath,stmaryrd}
\usepackage{mathptmx}
 \DeclareSymbolFont{letters}{OML}{txmi}{m}{it}
\usepackage[all]{xy}

\title{Towards a Formalization of Budgets}
\author{%
  Jan A.~Bergstra$^{1}$ \and
  Sanne Nolst Trenit\'{e}$^{2}$ \and
  Mark B.~van~der~Zwaag$^{1}$\\
\\
  {\small
	  ${}^1$Section Software Engineering,
	  Informatics Institute,
	  University of Amsterdam}\\
	{\small ${}^2$Faculty of Science,
	  University of Amsterdam
	}\\
	{\small Email: \{janb,sanne,mbz\}@science.uva.nl
	}
}
\date{}
\newcommand{\defeq}{\ensuremath{\stackrel{\text{\tiny{def}}}{=}}}
\newcommand{\Attr}{\ensuremath{\mathit{A}}}
\newcommand{\et}{\ensuremath{\varepsilon}} 
\newcommand{\nt}{\ensuremath{\delta}}
\newcommand{\conjc}{\ensuremath{\varobar}}
\newcommand{\ztest}[1]{\ensuremath{\gamma(#1)}}
\newcommand{\abs}[1]{\ensuremath{|{#1}|}}

\newcommand{\BudA}{\ensuremath{\mathit{A}}}
\newcommand{\BudB}{\ensuremath{\mathit{B}}}
\newcommand{\BudC}{\ensuremath{\mathit{C}}}
\newcommand{\BudJ}{\ensuremath{\mathit{J}}}
\newcommand{\STAFF}{\ensuremath{\text{STAFF}}}
\newcommand{\SES}{\ensuremath{\mathrm{SES}}}
\newcommand{\JES}{\ensuremath{\mathrm{JES}}}
\newcommand{\PM}{\ensuremath{\mathrm{PM}}}
\newcommand{\NDG}{\ensuremath{\mathrm{NDG}}}
\newcommand{\NEC}{\ensuremath{\mathrm{NEC}}}
\newcommand{\ESC}{\ensuremath{\mathrm{ESC}}}
\newcommand{\ECC}{\ensuremath{\mathrm{ECC}}}
\newcommand{\DGC}{\ensuremath{\mathrm{DGC}}}
\newcommand{\cpdg}{\ensuremath{\mathit{cpdg}}}
\newcommand{\escf}{\ensuremath{\mathit{escf}}}
\newcommand{\cpec}{\ensuremath{\mathit{cpec}}}
\newcommand{\bbp}{\ensuremath{\mathit{bbpp}}}
\newcommand{\ndg}{\ensuremath{\mathit{ndg}}}
\newcommand{\nec}{\ensuremath{\mathit{nec}}}
\newcommand{\af}{\ensuremath{\mathit{k}}}

\newcommand{\sslt}{\ensuremath{\mathit{sslt}}}
\newcommand{\sset}{\ensuremath{\mathit{sset}}}
\newcommand{\ssot}{\ensuremath{\mathit{ssot}}}
\newcommand{\pmt}{\ensuremath{\mathit{pmt}}}
\newcommand{\lpf}{\ensuremath{\mathit{lpf}}}
\newcommand{\sscph}{\ensuremath{\mathit{sscph}}}
\newcommand{\jscph}{\ensuremath{\mathit{jscph}}}
\newcommand{\jsst}{\ensuremath{\mathit{jsst}}}
\newcommand{\sspst}{\ensuremath{\mathit{sspst}}}
\newcommand{\jspst}{\ensuremath{\mathit{jspst}}}
\newcommand{\JSH}{\ensuremath{\mathrm{JSH}}}
\newcommand{\SSH}{\ensuremath{\mathrm{SSH}}}
\newcommand{\inc}{\ensuremath{\mathit{in}}}

\begin{document}

\maketitle
\begin{abstract}
We go into the need for, and the requirements on, 
a formal theory of budgets.
We present a simple algebraic theory of 
rational budgets, i.e., budgets in which amounts
of money are specified by functions on the rational numbers.
This theory is based on the tuplix calculus.
We go into the importance of using
totalized models for the rational numbers.
We present a case study
on the educational budget of
a university department offering
master programs.
\end{abstract}

\section{Introduction}
The process of budget design and financial accounting
is becoming increasingly specialized and exclusive. 
Unfortunately, the need for an underlying theory
seems to be unrecognized.
Economic theories of finance do not provide the
simple insights needed for
managing small-scale operations.
We are currently witnessing
the following developments.

\begin{enumerate}

\item 
Financial work takes place in a context of
complex IT support systems, which are often
poorly documented from a user perspective.
Documentation is typically
limited to user manuals, and does not give
conceptual descriptions of the underlying budget theory
and financial theories.

\item 
Financial competence is easily confused with the ability to 
operate
certain financial systems.
Because these systems are increasingly complex,
the competence to use them is becoming scarce and requires
training and experience.
Nevertheless that competence need not imply any deeper
awareness of the variation of business logics that may or
may not be served with a given system.

\item 
Financial planning is at the basis of many complex organizational
transformations.
Its logic is intimately connected with novel structural changes
such as outsourcing, insourcing, backsourcing and offshoring.
Organizational changes are often correlated with
changes in budget logic.

\end{enumerate}
In this situation, we find it worthwhile to
explore the applicability of modeling techniques developed
in the fields of information science and software engineering.
Unlike software architecture,
financial architecture seems to be a subject
to which relatively little attention is paid.
It is worth an effort to apply
system description techniques from computer science 
to financial systems and to facilitate
systematic and correct reasoning about them.
We believe that
financial architecture can profit from the same
development strategy as software architecture by making
use of a basis of design patterns and by developing
very clear modularization techniques.

In the other direction,
we imagine a formalization of budgets to be a helpful ingredient 
for the development of sourcing theory
(see for instance~\cite{D05} and~\cite{RD04}).
Sourcing theory requires the presence of so-called business cases
for insourcer and outsourcer
to be available and scrutinized before any deal is made.
After outsourcing has been executed,
services are expected to be
delivered in accordance with an SLA
(Service Level Agreement). 
Budget information is an essential part of any SLA.
No complete theory of sourcing is possible without
some theory of SLAs and their underlying budgets.
SLAs constitute a relatively new topic in computing and their
meta-theory is still in an initial stage.
We expect that SLAs, or similar entities existing 
within a further evolved terminology,
will become a central cornerstone of the emerging 
theory of service-oriented computing.

Overview.
In this article we define a simple algebraic theory of 
\emph{rational budgets}, that is, budgets in which amounts
of money are specified by functions on the rational numbers.
For this theory we borrow from our experience
with process algebras (modeling the behavior of
computer systems) and abstract data types.
In Section~\ref{sec:rat} we go into the use
of rational functions, and in particular
on the importance of
\emph{totalized\/} models for the rational numbers 
when defining them as an abstract data type.
In Section~\ref{sec:bl}
we reflect on the formalization of budgets,
and this is followed by a simple theory of budgets in 
Section~\ref{sec:ba}.
We use the so-called tuplix calculus~\cite{TC} to model budgets.
In Section~\ref{sec:casestudy}
we present a case study on the educational budget of
a university department offering master programs.

\section{Rational Numbers}\label{sec:rat}
It is a common misunderstanding that because budget figures
are to be understood as measuring quantities of money expressed
in terms of known currencies all semantic problems will disappear. 
The issue is comparable to a program notation designed for
programming computations on natural numbers. 
In spite of the seemingly clear mathematical basis of the 
program notation,
in the absence of a formally specified semantics of the program 
notation at hand, nothing much can be said about what 
transformations on natural numbers a specific program denotes.

For the definition of a budget the data type of rational
numbers is considered of central importance.
All financial quantities will me measured in exact rational 
numbers.
If the question arises what exactly are the rational numbers
we refer to~\cite{BT05} which provides a novel and concise
initial algebra specification of this classical mathematical
structure. 
We hold that division plays an important role in budgeting,
because of the need to distribute expected costs over a number
of expected users. 
If ten users will make mutually comparable use of a single shared 
service each of them will be expected to pay 10\% of its costs 
unless more specific information is available.
Interestingly, division seems 
not to feature in accounting and bookkeeping. 
In~\cite{BT05}, division is given a first class status with
an operator symbol reserved for it, very much 
like addition, multiplication and subtraction.
Moreover we will insist, against conventional practice,
that division is a total operator.
This leads to `meadows' of rational numbers:
a meadow is the well-know algebraic structure `field'
with a total operator for division, 
so that division by zero produces some value in 
the domain of the field. 
In a zero-totalized field division is made 
total by choosing zero as the result of 
division by zero (and, for example, in 
  a 47-totalized field one has chosen 47 to represent the result 
  of all divisions by zero).

The relevance for our theory of budgets is this:
budgets will contain expressions for
rational functions rather than 
`closed' figures in full (fixed point) precision. 
From the conventional perspective on rational 
numbers, these functions may be undefined for
certain input values, namely, for values leading
to devision by zero.
In general, it will be far from
trivial to decide whether functions
are always defined, and, if not, to establish
the values for which it is.
On the other side, the meadow of rational numbers
constitutes a total algebra with 
trivial type-checking properties,
that provides us with a clear meaning of expressions.
This will be of the highest importance to our endeavor,
because just like in the 
case of specifying computer programs, there will be 
no way to avoid explicit syntax and 
type-checking of expressions.

A down-side of working with zero-totalized fields is that some 
calculations will produce useless results.
In most cases the occurrence of division by zero in the course
of a calculation still 
indicates the presence of an error somewhere and error detection 
techniques will be needed.
Nevertheless the meta-theory of this form of error detection is
considered far simpler than the meta-theory of partial algebras,
thus creating a trade-off to the advantage of the 
use of calculation in zero-totalized fields.

A small digression: 
one might wonder why the issue to define division by zero
is so easily avoided in school mathematics and its academic 
sequel.
The answer is that in mathematics most specialists make no 
distinction between syntax and semantics.
No syntactic expression is entitled to any 
attention on the sole grounds of its formal existence.
If syntax is used its use follows the development of semantics
and there always is an intended meaning. 
Realistically, the question `what is the intended meaning of that 
piece of syntax' cannot be even posed. 
A reluctance to separate syntax and 
semantics may become a weakness when concepts need to 
be defined which are viewed as constructs of a syntactic nature 
consisting of parts rooted in classical mathematics.

\section{Theory and Practice of Budgets}\label{sec:bl}
A crucial point in the design of a formal theory of budgets
is the following separation of concerns.
Our task is to give a
conceptual, mathematical definition of budgets.
This definition (of what a budget \emph{is})
should be as much as possible 
independent of how and why budgets are \emph{used}.
A budget will not be assigned a 
behavior of its own a priori.
One wants to avoid definitions like `natural numbers are a
very practical concept that 
has been in use since the need arose to count sheep, for which 
natural numbers turn out to work very well fortunately.'
Clearly budgets are artifacts of a human origin, and, as in the 
case of natural numbers or data bases,
their use is independent of the artifact at hand. 

We believe that a formal theory of budgets
is essential to the analysis and improvement of
their use.
Some examples:

(1) 
The practice of budget design can be compared
to the practice of computer programming: in general 
programmers have no means available to know in advance what 
computers will do with their writings.
This uncertainty is mainly due to a lacking 
theoretical basis but equally to a common ethos which acts against 
the use of that theoretical basis even if it happens to exist and
if it might be readily available at reasonable costs.
In computer programming, testing is the main though unconvincing 
tool to fight this form of uncertainty. 
In budget design the concept 
of testing is significantly harder to imagine, however.
Budgets seem to be submitted to a number of static checks only.
Then after their use some form of evaluation and assessment may 
produce new guidelines (design rules) for budgeting and new 
budgets will be matched with these new design rules as well.
Budget testing comparable to dynamic testing of control code will
require a simulation environment.
That environment is quite specific to an 
organization and most organizations have no such tool available to 
them at the time of this writing.
Clearly, budget simulation is a hopeless endeavor without
a formal theory of budgets.

(2) 
Irrespective of the objectives a budget designer has in mind 
when writing a budget, it will evolve through a life-cycle.
An organization may prescribe this life-cycle to its budget 
designers in very much the same way
as a software life-cycle may play a normative role in a
computer software production factory.
Like a machine control code can be active (running, 
executing) a budget can be used to control events within an 
organization.
We would hope that when 
we start with a clear formal definition, we may be able to
explain (at least in principle) how that 
form of control might work.
Also, as budgets are often very context-specific,
they may be compared to dedicated computer programs. Renewing a 
budget on an annual basis may be compared to computer software 
maintenance
(although that comparison may well underestimate the degree of 
innovation that a new budget requires).

(3) 
When hard-pressed to qualify the writing of a budget,
the following viewpoint might be reasonable: budgets are 
proposed in a context in which their proposal is best viewed as a 
move from the side of its author in a game which is implicitly
present in the mentioned context.
Let us assume that a budget is designed for an activity
called $A$.
After having been designed and worked out 
in detail, the distribution of a budget
can be considered a move in a game.
By introducing a budget 
proposal the decision-making process is somehow influenced.
One may assume that this process will eventually lead to a 
validated budget for activity $A$.
The budget proposal may impact the style of budget design 
which will be adopted for $A$ and for similar activities.
The intriguing observation is that when time has come to write 
budgets for $A$, many different and competing budget proposals may be 
simultaneously put forward. Thus at budget design time there is no 
such thing as `the budget' in very much the same way as a computer 
program under construction leaves open many degrees of freedom.
It may be fruitful to experiment with writing quite different budgets
for the financial control of a single activity $A$. 
Again, such analysis of the practice
of budgeting should start with a formal theory of budgets.

\section{A Simple Budget Algebra}\label{sec:ba}
We present a simple algebra for rational budgets.
This algebra is an application of
the so-called tuplix calculus~\cite{TC}. 
A tuplix (plural: tuplices) is
a datastructure that collects
attribute-value pairs. The tuplix calculus
provides signature and axioms for various
operations on budgets, including 
a means to express constraints on budgets,
the composition of budgets, and encapsulation.
So, budgets will be given by means of tuplix expressions. 
The design of user-friendly syntax for budget expressions is
outside the scope of this paper.
However, the definition of budgets 
as tuplices provides a rudimentary syntax which will 
suffice for explanatory purposes.

\subsection{Entries and Tests}
The basic building blocks of
budgets are \emph{entries} and \emph{tests}.
An entry is an attribute-value pair of the form
\[ a(p)
\]
where $a$ is an attribute from a given set \Attr\
of attribute symbols, and $p$ is a data term.
For the values we use
the data type of rational numbers,
which we assume to be given by a zero-totalized field
as explained in Section~\ref{sec:rat}
(so $p/q$ is always defined).
An entry represents a payment: the attribute is
used in the communication between payer and payee,
and describes or identifies a transaction;
we refer to the attribute as the \emph{channel}
of the transaction, and shall also say that
the payment occurs \emph{along} the channel. 
The term $p$ represents the amount of money involved.
An entry $a(p)$ with $p>0$ stands for
an obligation to pay amount $p$ along channel $a$.
If $p<0$, the entry stands for the expected receipt
of amount $p$ along $a$.

A \emph{zero test} is a term of the form
\[ \ztest{p}
\]
for amount $p$.
It acts as a conditional:
if the argument $p$ equals zero,
then the test is void and disappears from compositions;
if the test is not equal to zero, it nullifies
any composition containing it.
Observe that an equality test $p=q$ can be
expressed as \ztest{p-q}.

\subsection{Budget Composition}
We define a budget as a (conjunctive)
composition of entries and zero tests.
This composition is commutative and associative:
\begin{align}
\label{ax:comm} x\conjc y &= y \conjc x,\\
\label{ax:ass} (x\conjc y)\conjc z &= x\conjc(y \conjc z).
\end{align}
There are two constants for budgets:
the \emph{empty budget}, notation \et,
and the \emph{null budget}, notation \nt.
The empty budget stands for the absence
of entries or tests, and the null budget
is used to model an erroneous situation which
nullifies the entire composition containing it.
Axioms:
\begin{align}
\label{ax:eb} x\conjc \et &= x,\\
\label{ax:ib} x\conjc \nt &= \nt.
\end{align}

Entries with the same attribute
can be combined: 
\begin{align}
\label{ax:acc} a(u)\conjc a(v) &= a(u+v).
\end{align}
Note in particular that the composition
of a payment $a(p)$ and the receipt $a(-p)$
can be reduced to $a(0)$.
We shall see below that \emph{encapsulation}
both enforces and hides such synchronizations.

For the axiomatization of the zero tests, we use 
the property that in 
the zero-totalized field for the rational numbers,
the division $p/p$ yields zero only if 
$p$ is equal to zero; otherwise it yields $1$.
Axioms:
\begin{align}
\ztest u &= \ztest{u/u},\\
\ztest 0 &= \et,\\
\ztest 1 &= \nt.
\end{align}
For reasoning about budgets with open data terms, 
we add the following two axioms:
\begin{align}
\label{ax:T9}\ztest{u}\conjc\ztest{v}&= \ztest{u/u+v/v},\\
\ztest{u-v}\conjc a(u)&= \ztest{u-v}\conjc a(v).
\end{align} 

\subsection{Encapsulation}
For set of attributes $H\subseteq\Attr$,
the operator $\partial_H(x)$
encapsulates all entries with attribute $a\in H$
occurring in $x$.
That is, if the
accumulation of quantities in entries with attribute $a$
equals zero,
the encapsulation on $a$ is considered successful
and the $a$-entries disappear; 
if the accumulation is not equal to zero, 
it yields the null budget \nt.
Axioms:
\begin{align}
\partial_H(\et) &= \et,\\
\partial_H(\nt) &= \nt,\\
\partial_H(\ztest{u}) &= \ztest{u},\\
\label{ax:E4}\partial_H(a(u)) &=
  \begin{cases}
   \ztest{u} & \text{if } a\in H,\\
   a(u)      & \text{if } a\not\in H,
   \end{cases}\\
\label{ax:E5}\partial_H(x\conjc\partial_H(y)) &=
   \partial_H(x)\conjc\partial_H(y).
\end{align}
We further adopt the identities
\[ \partial_{H\cup H'}(x)=\partial_H\circ\partial_{H'}(x)\quad
    \text{and}\quad
    \partial_\emptyset(x)=x.
\]

Example.
Consider budget
\[ P \defeq a(-30)\conjc b(10)\conjc b(20).
\]
This budget specifies the expected receipt of 
amount 30 along channel $a$ and payments
of amount 10 and of amount 20 along $b$.
We compose it with budget
\[ Q \defeq b(-30)\conjc c(30),
\]
which specifies that amount $30$
is received along $b$ and sent along channel $c$.
We see that 
the payments of $P$ will match the receipt of $Q$ on 
channel $b$, so that encapsulation of $b$ will hide
these entries:
\[ \partial_{\{b\}}(P\conjc Q) = a(-30)\conjc c(30).
\]
To derive this, first derive that
\[ \partial_{\{b\}}(a(-30)\conjc c(30))= a(-30)\conjc c(30)
\]
using axioms \eqref{ax:E4} and \eqref{ax:E5}.
Then:
\begin{align*}
\partial_{\{b\}}(P\conjc Q) 
  &=\partial_{\{b\}}(a(-30)\conjc b(10)\conjc b(20)\conjc b(-30)
\conjc c(30))\\
  &=\partial_{\{b\}}(b(0)\conjc a(-30)\conjc c(30))\\
  &=\partial_{\{b\}}(b(0)\conjc \partial_{\{b\}}(a(-30)\conjc c
(30)))\\
  &=\partial_{\{b\}}(b(0))\conjc \partial_{\{b\}}(a(-30)\conjc c
(30))\\
  &=\ztest{0}\conjc a(-30)\conjc c(30)\\
  &=a(-30)\conjc c(30).
\end{align*}

Another example.
When computing the encapsulation
of more than one channel,
we split the encapsulation up, and compute them one by one.
Recall that we defined
\[ \partial_{H\cup H'}(x)=\partial_H\circ\partial_{H'}(x).
\]
For example, we derive
\[ \partial_{\{a,b\}}(a(0)\conjc b(0)) 
   = \partial_{\{a\}}\circ\partial_{\{b\}}(a(0)\conjc b(0))
   = \partial_{\{a\}}(a(0))
   = \et.
\]

\subsection{Constraints}
In the case study
in this article, we assume that
an absolute operator $\abs{\_}$
(defined by $\abs p = p$ if $p\geq 0$, and $\abs{p} = -p$ 
otherwise)
is part of the signature for rationals.
With this operator 
we can express inequalities:
\[ \ztest{\abs{q-p}-(q-p)}
\]
expresses the test $p\leq q$. 
For inequality tests we shall then simply write
$\ztest{p\leq q}$.
We sometimes write \ztest{p=q} for \ztest{p-q}.

For example, we may design a budget under the constraint 
\[ \phi \defeq p\leq q,
\]
and we then compose the budget with the 
test $\ztest\phi$.
For the composition under multiple
constraints, say $\phi$ and $\psi$,
we may use the notation
\[ \ztest{\phi\land\psi} \defeq \ztest{\phi/\phi + \psi/\psi} 
    \stackrel{\eqref{ax:T9}}{=} \ztest{\phi}\conjc\ztest\psi.
\]

\section{Case Study: MSc Program Budgets}\label{sec:casestudy}
We consider a university department that maintains
the three MSc programs A, B and C\@.
Each program offers a 1-year, 
60 EC\footnote{An EC is a unit of student activity/learning 
  outcome in the European Credit Transfer System.
  One EC stands for 28 hours of work.}
curriculum.
These programs need a new budget because of changes concerning
budget guidelines, financial reporting, risk management and
business accounting.
Below we develop budgets for the programs.
Having fixed these budgets, 
the three program managers should negotiate the setting of
certain variables. 
Having done that the program managers are free to develop their
programs within the constraints of the budget. 
This process may be viewed as a game aimed at
the design of a single budget where coalitions try to
get things their way by imposing preferred variable settings
on other participants.
In our case, the program managers will not hesitate to get 
money their way at the expense of the other programs or
to prove other programs financially unsound,
should they find possibilities to do so in the new 
system.\footnote{Mark Burgess (see for instance~\cite{Bu05})
  advocates the mechanism of autonomous agents making
  promises which constrain their actions on a voluntary
  basis only.
  A budget proposal might be viewed as a promise conditional under
  the counter promise by other parties that they will go along
  with it. 
  Finding protocols for distributed budget design in a context
  of voluntary cooperation is an interesting challenge 
  for further research. 
}

\subsection{Generic Structure of an MSc Program}
Each of the three programs has the following structure: 
\begin{enumerate}
\item An introduction week providing general information. 
\item 4 courses of 10 EC each. A course consists of 
  300 working hours composed from these ingredients: 
  \begin{itemize}
  \item Between 40 and 160 hours of teaching by senior staff. 
  \item Working group meetings supervised by junior staff. 
  \item Unsupervised student team meetings. 
  \item Unsupervised individual experimental work. 
  \item Unsupervised individual homework. 
   \item Participation in an examination.
   \end{itemize}
\item 2 projects of 10 EC each.
 A project is supervised in one of the following ways:
  \begin{itemize}
  \item Between 5 and 10 hours of internal
  senior staff supervision.
  \item At most 20 hours of junior staff supervision.
  \item At most 5 hours of external staff supervision
  (performed outside the institution).
  \end{itemize}
  A project ends with a 30-minute presentation 
  (with at least two senior staff members present).
\item A formal final degree ceremony.
\end{enumerate}

Each program offers at least two mandatory
courses for its own students, and
may offer a number of optional courses
that can also be followed by students from the other programs.
An expensive way to implement this is to offer six dedicated
courses in the program and to make two of these compulsory while 
leaving the students the option to choose two from the other
four courses.
A reason to use this kind of planning may be to let research staff
lecture about their advanced topics in order to recruit
future PhD students.
Another reason might be to make sure that the entire student
population acquires a wide body of knowledge representative of the
field as a whole while accepting that each individual student has
acquired knowledge in a more limited scope. 
A much cheaper option is to offer only the 2 mandatory courses
and to ask the students to take electives from 
courses offered by other programs.
Reason for doing so may be a lack of staff or financial resources.
Another reason might be the intention to educate
a homogeneous group of experts who will be able to cooperate
effectively in forthcoming projects.

\subsection{Joint Budget}
In this case study we specify four budgets:
budgets \BudA, \BudB, and \BudC, for the 
respective programs A, B, and C,
and one joint budget \BudJ. 
We start with the joint budget.

All income of the programs from external
sources is specified in
the joint budget, and these incoming
amounts are received via channel $\inc$.
The task of the joint budget
is to specify the distribution
of the income between
the three programs and shared costs.
Payments from the joint budget to the
individual program budgets
run via the respective 
channels $a$, $b$, and $c$.
The only shared costs are the payments
to the so-called educational
service center 
 (student consulting, time tabling, lecture hall
  reservation, facility management, administration);
these payments are done via channel $e$.
Picture:
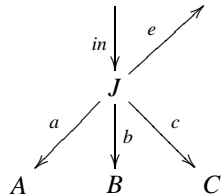
\[
\xymatrix{
 & \ar[d]_\inc &\\
 & \BudJ \ar[ur]^e \ar[dl]_a \ar[d]^b\ar[dr]^c\\
 \BudA & \BudB & \BudC
}
\]
When we consider the composition of the four budgets, 
payments along the
channels $a$, $b$ and $c$ are considered to be internal.
The channels \inc\ and $e$ are external; they `link'
to parties for which we do not have the budgets.

Notation.
The letter $X$ ranges over
$A$, $B$, $C$ (denoting the programs).
We use lowercase italics for variable names,
and uppercase roman for abbreviations.

\begin{figure}
\begin{center}
\fbox{
\begin{minipage}{11cm}
\begin{tabular}{rp{8.5cm}}
\multicolumn{2}{l}{\textbf{Set by external authority:}}\\
\cpec & The \emph{compensation per EC}.
 Amount obtained when awarding one EC.\\
\cpdg & The \emph{compensation per degree}.
 Amount obtained when awarding one degree.\\
\escf & The \emph{educational service
 center fraction} (value between 0 and 1)
 of the overall income
 to be transferred to the educational service
 center.\\[1ex]
\multicolumn{2}{l}{\textbf{Set by measurement:}}\\
$X{:}\nec$ & Total number of EC awarded
 in courses offered by program X.\\
$X{:}\ndg$ & Number of degrees awarded
 in program X.\\[1ex]
\multicolumn{2}{l}{\textbf{Set by budget designer/negotiation:}}\\
\bbp & A fixed amount
 that serves as the \emph{basic budget per program}.
 This amount is equal for each program and is used
 to pay for the program manager, various committee tasks,
 marketing and communication.\\
\af & Fraction (value between 0 and 1)
 of the \bbp\ which is taken from the part of the 
 overall income stemming from degree compensation.
 The remaining fraction $1-\af$ 
 of the \bbp\ is taken from the overall EC compensation.
\end{tabular}
\end{minipage}
}
\end{center}
\caption{Variables used in budget \BudJ}
\label{fig:varsJ}
\end{figure}

Variables. The variables used
in the specification of the joint
budget \BudJ\ are listed in
Figure~\ref{fig:varsJ}.
The values for the variables 
$X{:}\nec$ and 
$X{:}\ndg$ are determined by 
measurement and monitoring
(in practice one may take last years numbers instead). 
The values for the variables \bbp\ and \af\
are determined by negotiation
between the program managers.
We shall return to the consequences
of the setting of these variables.

Income. The income, received via channel \inc,
consists of two parts.

First, 
there is the overall EC compensation (\ECC),
defined as the total number of
EC credits awarded in the three
programs, multiplied by the compensation
per credit:
\begin{align*}
 \NEC &\defeq \sum_X X{:}\nec,\\
 \ECC &\defeq \NEC\cdot\cpec.
\end{align*}
Similarly,
there is the overall degree compensation (\DGC),
defined as the total number of
degrees awarded in the three
programs, multiplied by the compensation
per degree:
\begin{align*}
 \NDG &\defeq \sum_X X{:}\ndg,\\
 \DGC &\defeq \NDG\cdot\cpdg.
\end{align*}

Expenses.
The educational service center (ESC) takes care of
all data base handling, time tabling, logistics,
communication and marketing, help desks of various kinds, 
international
relations and formal ceremony management.
There is a joint payment to the ESC, consisting
of the fraction \escf\/ of the overall income:
\[ \ESC\defeq \escf\cdot (\ECC+\DGC).
\]

The remainder $(1-\escf)\cdot(\ECC+\DGC)$
of the overall income
is distributed among the three programs.

First, each program receives the amount \bbp.
Of course, this amount cannot be more
than one third of the available money,
so we adopt the constraint
\begin{equation*}
\phi_1\defeq \bbp\leq (1/3)\cdot (1-\escf)\cdot(\ECC+\DGC).
\end{equation*}
Furthermore, we require that fraction $\af$ of
the \bbp\ is taken from
the overall degree compensation,
and the remaining part $(1-\af)$
of the \bbp\ is taken from the
overall EC compensation,
leading to the following constraints:
\begin{align*}
\phi_2 &\defeq \af \leq (\DGC\cdot (1-\escf))/(3\cdot\bbp),\\
\phi_3 &\defeq (1 - \af) \leq 
 (\ECC\cdot (1-\escf))/(3\cdot\bbp).
\end{align*}
Apart from the fixed amount \bbp,
that provides each program with a financial basis independent 
of its own student numbers (assuming that the other programs 
have sufficient numbers of students),
each program gets a share 
of the remaining part of the
overall EC and degree compensation.
These shares are
proportional to the contribution that 
the program has in the overall compensation. 
The remaining part of the
degree compensation, after 
subtraction of the expenses on \ESC\ and \bbp,
is
\[ \DGC\cdot(1-\escf)-3\cdot\af\cdot\bbp,
\]
and the share of this amount
awarded to program X is $X{:}\ndg/\NDG$.
So we define
\[
 X{:}\DGC\defeq
  (\DGC\cdot(1-\escf)-3\cdot\af\cdot\bbp)
  \cdot 
  X{:}\ndg/\NDG.
\]
Similarly, we define
\[
 X{:}\ECC\defeq
  (\ECC\cdot(1-\escf)-3\cdot(1-\af)\cdot\bbp)
  \cdot
  X{:}\nec/\NEC.
\]
Each program X receives from
the joint budget the amount
\[ X{:}\STAFF\defeq \bbp+X{:}\DGC+X{:}\ECC.
\]

Budget.
Putting everything together,
the joint budget \BudJ\ is defined by
\begin{align*}
 \BudJ
 \defeq {}&
 \ztest\phi\conjc \inc(-\ECC)\conjc \inc(-\DGC)\conjc
   e(\ESC)\conjc{}\\
 &\quad a(A{:}\STAFF)\conjc 
   b(B{:}\STAFF)\conjc
   c(C{:}\STAFF),
\end{align*}
where
\[ \phi\defeq\phi_1\land\phi_2\land\phi_3.
\]

\paragraph{Notes.}
The joint budget has been designed with the following
properties in mind.
\begin{itemize}

\item 
By taking \bbp\ low (or simply zero) each budget gets as much as
possible resources proportional to its production in EC and in
degrees.
By taking \bbp\ higher each program budget is provided with a
minimum funding with the effect that each program gets less return
on investment for a single EC or degree.

For example, a program with relatively few students
may strive for a significant fixed budget basis \bbp\
(maybe even 
$\bbp = (1/ 3)\cdot (1 - \escf)\cdot ( \DGC + \ECC)$
in an extreme case).

\item
By taking \af\ low (or simply zero) a maximal reward is
provided for programs with a high yield in terms of degrees.
By taking \af\ high (or simply 1) a maximal reward is given to
programs that get as many as possible ECs to students irrespective
of their program and irrespective of whether or not they will
complete their degree.

For example, again for a program with relatively few students:
choose fraction \af\ as close as possible to 1
(for all values of \bbp, 
\[ \af = (1/3)\cdot (\DGC \cdot (1 - \escf))/ \bbp
\]
seems to be a reasonable choice).
By taking \af\ as close as possible to 1, 
the program will profit the most from students from the other 
programs following its courses. 

\item
The following constraints need not be imposed as they follow from
the defining equations, that is by adding these constraints
the meaning of the budget will not change:
\begin{align*}
\sum_X X{:}\DGC &= \DGC \cdot (1 - \escf )-3\cdot\af\cdot\bbp,\\
\sum_X X{:}\ECC &= \ECC \cdot (1 - \escf )-3\cdot(1-\af)\cdot
\bbp. 
\end{align*}

\end{itemize}

\subsection{Budgets per Program}
In the individual program budgets,
the amount received from the joint budget is
spent on the following costs:
\begin{itemize}
\item Senior Educational Staff (\SES).
Compensation for educational working hours by senior staff.
\item Junior Educational Staff (\JES).
Compensation for educational working hours by junior staff.
\item Program Management (\PM).
Compensation for all forms of program management
performed by educational staff.
\end{itemize}

Variables.
The variables used in the
program budgets are listed in
Figure~\ref{fig:varsX}.
Within the constraints set by the educational budget,
a program manager can vary the values for 
the second set of variables.
Needless to say this leads to 
a combinatorial explosion of options. 
Setting these variables low implies a sound budget 
but introduces risk with student
success rates, with student satisfaction monitoring and
periodically with external quality control authorities.
Most importantly, however, setting the other variables very
low will cause senior staff to complain about unrealistic
requirements and workloads.

\begin{figure}
\begin{center}
\fbox{
\begin{minipage}{11cm}
\begin{tabular}{rp{8.5cm}}
\multicolumn{2}{l}{\textbf{Set by external authority:}}\\
\sscph & Senior staff marginal integral cost per hour.\\
\jscph & Junior staff marginal integral cost per hour.
\\[1ex]
\multicolumn{2}{l}{\textbf{Set by program manager:}}\\
$X{:}\lpf$ &
 Lecture preparation factor (the number of hours used to
 prepare one hour of lecturing).\\
$X{:}\sset$ & Senior staff examination time:
  time needed to set and mark an exam.\\
$X{:}\sspst$ & Senior staff project supervision time
 (number of hours spent
  by senior staff supervising a single student project).\\
$X{:}\jspst$ & Junior staff project supervision time
 (number of hours spent by junior staff supervising a single
 student project) in addition to senior staff supervision.\\
$X{:}\ssot$ & Senior staff overhead in total (hours per year).\\
$X{:}\pmt$ & Program management time (hours per year spent
  by program manager).\\
$X{:}C{:}\sslt$ & Senior staff lecturing time
  (number of hours) for course C.\\
$X{:}C{:}\jsst$ & Junior staff supervision time
 (number of hours) for course C.
\end{tabular}
\end{minipage}
}
\end{center}
\caption{Variables used in program budgets}
\label{fig:varsX}
\end{figure}

Budgets.
Define
the senior staff working hours (\SSH)
and junior staff working hours (\JSH)
as follows, with $C$
ranging over the set $C_X$ of courses
offered by program $X$.
\begin{align*}
X{:}\SSH &\defeq 
 \sum_C (X{:}C{:}\sslt \cdot (1 + X{:}\lpf ) + X{:}\sset) + 
  X{:}\ndg \cdot 2 \cdot X{:}\sspst\\
X{:}\JSH &\defeq \sum_C (X{:}C{:}\jsst) + 
             X{:}\ndg \cdot 2\cdot X{:}\jspst
\end{align*}
The factor 2 in the summands for project supervision
stems from the fact that each student does two projects.
Note that project supervision generates compensation only when
students obtain their degree.
Failed projects or projects for students failing elsewhere in 
the program will not generate financial resources. 

The expenses for senior and junior
staff are found by multiplying
their hours by their
respective marginal integral costs per hour:
\begin{align*}
X{:}\SES &\defeq X{:}\SSH \cdot \sscph,\\
X{:}\JES &\defeq X{:}\JSH \cdot \jscph.
\end{align*}
Finally, the
staff costs for program management are given by
\[ X{:}\PM \defeq (X{:}\ssot + X{:}\pmt) \cdot \sscph.
\]

Budgets.
\begin{align*}
\BudA &\defeq a(-A{:}\SES-A{:}\JES-A{:}\PM)\\
\BudB &\defeq b(-B{:}\SES-B{:}\JES-B{:}\PM)\\
\BudC &\defeq c(-C{:}\SES-C{:}\JES-C{:}\PM)
\end{align*}

\paragraph{Notes.}
\begin{enumerate}

\item
Both senior staff members and junior staff members may
spend two kinds of hours:
regular office hours and spare time hours.
The second kind of work is unpaid.
No one will be ever forced to work without compensation but a
culture may exist where this is done on a regular basis.
It is quite common to perform unpaid
research work outside regular hours.
This is possible for teaching just as well.
Nevertheless there are some constraints.
All formal teaching, all introductory activity,
all examinations and all senior and junior staff supervision
must take place within office hours
and the cost of this staff time is given by fixed rates per hour. 
Office hours are either classified as educational office hours or
as research office hours or as unclassified office hours.
If educational activity is performed in research office hours it 
need not be paid from the educational budget but it will be paid
from the research budget instead.
This mechanism allows a research group to subsidize its teaching
activities from its research budget.
That subsidy may be justified in the case educating a small group 
of students brings about a few PhD students who might be very hard
to find otherwise. 

\item
In these budgets per program a formidable amount of freedom exists
because all variables determining the amount of junior and senior
staff time spent on courses and projects can me defined 
specifically for each of the programs.
If projects are very close to staff research they may be 
supervised in (seemingly) less time because additional unspecified
research time is used for the supervision as well while those hours
are not paid from this educational budget.
If for instance program A has rather few students in comparison
with the other two the following variable 
settings (or rather suggestions
for setting variables) can be helpful:
	\begin{enumerate}
	\item Set low senior staff hours for project supervision,
	and compensate
	that setting with higher junior staff hours and with time
	paid for from
	the research budget which is not viewed as a part of these
	budgets.
	\item Reduce the number of contact hours in course lecturing
	while requiring 
	a significant amount of autonomous work from the students
	and setting difficult exam papers.
	This strategy may backfire when students from the 
	other programs are supposed to attend the same lectures, 
	however.
	\end{enumerate}
	
\end{enumerate}

\subsection{Synchronization of the Budgets}
We have presented the four budgets
\BudA, \BudB, \BudC, and \BudJ.
The combined budget is the synchronization 
\[ \partial_{\{a,b,c\}}(\BudA\conjc\BudB\conjc\BudC\conjc\BudJ).
\]
We derive that it equals
\[
  \ztest{\phi\land\psi_A\land\psi_B\land\psi_C}
  \conjc
      \inc(-(\DGC+\ECC))\conjc e(\ESC)
\]
where
\[ \psi_X\defeq (X{:}\STAFF = X{:}\SES+X{:}\JES+X{:}\PM).
\]

Conclusion.
This example shows how a modular decomposition of
a budget can be designed.
The decomposition is valid under a number of conditions only.
If these conditions are not met,
further refinement of the budgets is needed.
That can be done by means of the same notation of course.

These conditions describe an abstraction level in the sense
that they rule out circumstances which may be of practical 
interest but which are considered an undesirable overhead at a
certain stage of design.

\subsection{Further Reflections}

\paragraph{Methods of Cost Measurement.}
Even if one observes the course of action when a particular 
program is run in all detail it is still difficult to make a
precise statement concerning its costs. 
A first difficulty is how to count or incorporate
free time hours made by staff members.
A second difficulty is how to decide if any official research
time is used for educational purposes.
A third issue is to determine the border between research 
and preparation for lectures and project supervision. 
A further complication for cost measurement based on
observations on how the work is actually done is this: 
if staff members are made aware of how their 
time investment is counted they may change their behavior.
Suppose a staff member often works additional unpaid hours at home
to get research done and then finds out that a counting
system has detected many hours spent on teaching within the
institution.
This may lead to the conclusion that the position will be
reclassified into a teaching position with a limited
research task only.
Then of course this staff member may be inclined to
interchange a number of
educational support activities
(marking exams, preparing lectures, etc)
with research activities that were done outside the office hours.
Thus a counting system should be stable in the sense that its 
introduction should not by itself influence staff behavior
in such a way that the results of counting are modified.
In order to obtain this form of stability staff members should be
given a variety of options for formally accounting their time. 

\paragraph{Budgets versus Costs and Planning.}
Given the combinatorial explosion of options for planning an MSc 
program in the formats given above it is an unreasonable request
to `offer the program in a cheaper form'
unless very clear goals are stated in advance.
The way in which a curriculum (that is, the listing of course 
titles and of project proposals)
can be offered and planned implies
that it is useless to suggest a
change on financial grounds unless a quite clear model of costs 
and revenues is available and unless a clear target in that model
has been set in advance.

\section{Conclusion}
We have motivated our interest in
a formal theory of budgets, and we
have proposed a simple algebraic
theory of budgets based on the tuplix calculus~\cite{TC}.
Quantities are expressed as functions on the
rational numbers, which we have modeled as a totalized 
field~\cite{BT05}.
As a case study, we modeled budgets and their
composition for a university department offering
master programs.
We have kept the theory simple, but
think of extensions such as operators for
choice (as present in the tuplix calculus),
binding of rational variables, a theory of interfaces
and hiding, etc.

As a preliminary conclusion we state that
budgets are amenable to formalization in the the data type 
tradition of theoretical computer science.
The example demonstrates that formalization
can be helpful to specify details which are likely to be missed in 
a less formal treatment and which are helpful for a proper 
understanding. 
At the same time, while working on the example, we have drawn the
conclusion that designing budgets in a modular fashion is not an
obvious matter and that many more cases studies will be needed to
obtain a stable and formalized structure theory of budgets.

\end{document}